\documentclass[twocolumn,english,prb,showpacs]{revtex4-1}
\usepackage[colorlinks=true,urlcolor=blue,citecolor=blue,linkcolor=blue]{hyperref}
\usepackage[T1]{fontenc}
\usepackage[latin9]{inputenc}
\usepackage{amssymb}
\usepackage{graphicx}
\usepackage{amsmath,color}
\usepackage{mathrsfs}
\usepackage{float}
\usepackage{indentfirst}
\usepackage{babel}
\usepackage[sort&compress]{natbib}
\usepackage{color}

\newcommand{\eqnref}[1]{Eq.~(\ref{#1})}

\begin{document}

\title{Entanglement as a resource in adiabatic quantum optimization}

\author{Bela Bauer}
\affiliation{Station Q, Microsoft Research, Santa Barbara, CA 93106-6105, USA}

\author{Lei Wang}
\affiliation{Theoretische Physik, ETH Zurich, 8093 Zurich, Switzerland}

\author{Iztok Pi\v{z}orn}
\affiliation{Theoretische Physik, ETH Zurich, 8093 Zurich, Switzerland}

\author{Matthias Troyer}
\affiliation{Theoretische Physik, ETH Zurich, 8093 Zurich, Switzerland}

\begin{abstract}
We explore the role of entanglement in adiabatic quantum optimization by performing approximate
simulations of the real-time evolution of a quantum system while limiting the amount of entanglement.
To classically simulate the time evolution of the
system with a limited amount of entanglement, we represent the quantum state using matrix-product states and projected
entangled-pair states.
We show that the probability of finding the ground state of an Ising spin glass on either a planar or non-planar
two-dimensional graph 
increases rapidly as the amount of entanglement in the state is increased. Furthermore,
we propose evolution in complex time as a way to improve simulated adiabatic evolution and
mimic the effects of thermal cooling of the quantum annealer.
\end{abstract}
\maketitle

\section{Introduction}

The concept of a quantum computer, originally proposed by Feynman in 1982,\cite{feynman1982} has stirred decades of both experimental and theoretical research. Quantum computers represent a fundamentally more powerful model of computation than classical Turing machines, and allow certain problems, such as factoring\cite{Shor1994} and simulation of quantum dynamics,\cite{feynman1982} to be solved exponentially faster than by the best known classical algorithm. The experimental control of quantum systems has now advanced to the point where a small quantum computer appears feasible. This has spurred a search for problems that a small-scale quantum device with a limited number of imperfectly controlled qubits can solve more efficiently than a classical computer.

A proposed approach that may perform very well under such circumstances is adiabatic quantum optimization,~\cite{Ray1989,Finnila1994,Kadowaki1998,arnab2008} a variant of adiabatic quantum computation,\cite{Farhi2000} tailored to solve a classical optimization problem by adiabatically tuning between a Hamiltonian whose ground state is easily prepared, and one that encodes the optimization problem to be solved.
To be specific, we consider the Ising spin glass, which we can describe by the Hamiltonian
\begin{equation} \label{eq:Ising}
  H_\mathrm{Ising} = -\sum_{i < j} J_{ij} s_i s_j,
\end{equation}
where $s_i = \pm 1$. Finding the ground state of this Ising spin glass on a non-planar graph is an NP-hard task,\cite{barahona1982} i.e. no algorithm is known that can solve hard instances in polynomial time. Local search algorithms, such as simulated annealing,\cite{kirkpatrick1983} often become trapped in local minima. It has been suggested that quantum annealing can exploit quantum tunneling effects to escape these local minima. To realize this, we represent the classical Ising spins by $S=1/2$ quantum spins, replacing each $s_i$ with a Pauli matrix $\sigma^z_i$. In the most common approach, one prepares all spins in the ground state of the initial Hamiltonian $H_X = -\Gamma \sum_i \sigma^x_i$. By adiabatically turning off the transverse field $\Gamma$ and increasing the Ising couplings $J_{ij}$, one can find the ground state of the classical model. The algorithm is thus to solve the time evolution
for the time-dependent Hamiltonian
\begin{equation} \label{eq:TFIM}
H(s) = s H_\mathrm{Ising} + (1-s) H_X,
\end{equation}
where $s=t/T \in [0,1]$ and $T$ is the total annealing time. Here, the evolution must be performed slowly enough to ensure adiabaticity against the minimal gap of \eqnref{eq:TFIM} for any $s$.

While this approach seems promising and has even been pursued in hardware,~\cite{Johnson2011} the potential of quantum annealing as a general-purpose
optimizer remains unclear and to date no speedup has been conclusively demonstrated in experiments on actual devices.\cite{ronnow2014}
Quantum annealing can also be simulated or mimicked in classical algorithms.\cite{troyer2014quantum} While the brute-force simulation of the quantum dynamics is prohibitively
expensive, many approximate classical approaches to quantum annealing have been explored, including quantum Monte Carlo simulations,\cite{santoro2002theory,Boixo:2014cg,heim2014quantum}
mean-field models,\cite{Smolin:2013vpa,shin2014,Wang:2013tga} and matrix-product states. \cite{banuls2006,crowley2014quantum}

It is a widely held belief that the power of quantum computers resides with their ability to compute using entangled states,\footnote{Exceptions from this are known for computation with mixed states\cite{Biham2004,lanyon2008}.} making quantum entanglement the key resource in quantum computing.
In this paper, we address the question of whether entanglement is a useful resource also in adiabatic quantum optimization, and explore a number of classical approaches to approximate quantum annealing. To assess the role of entanglement, we simulate quantum anealing using two different tensor network approaches, namely matrix-product states\cite{white1992,white1992-1,ostlund1995} (MPS) and projected entangled-pair states (PEPS).\cite{verstraete2004} These methods allow us to limit the amount of entanglement in the state at any point in the evolution by tuning their bond dimension.

We improve the reliability of approximate adiabatic evolution schemes by evolving in \emph{complex} time,\cite{troyer2014quantum} 
i.e. performing evolution not under $\exp (-i \delta t H)$, but instead $\exp( - (i+\epsilon) \delta t  H)$. For small $\epsilon$, where the evolution is close to real-time, this approach effectively reduces errors due to non-adiabaticity or approximations in the classical algorithm. This is in many ways reminiscient of coupling a quantum system to a cold heat bath.

\section{Mean-field dynamics  \label{sec:clustermf}}

In the absence of entanglement, the dynamics can be treated within a mean-field approach of product states by replacing for each spin the quantum
wavefunction $\psi_i \in \mathbb{C}^2$
by a classical variable $\vec{M}_i \equiv \psi_i^{\dagger} \vec{\sigma} \psi_i$, where $\vec{\sigma} = (\sigma^x, \sigma^y, \sigma^z)^T$ are
the Pauli matrices. The dynamics of the classical variable are governed by
  \begin{eqnarray}
  \frac{\partial \vec{M}_i}{\partial t}  = \vec{h}_i(t/T)
  \times \vec{M}_i \ ,
\end{eqnarray}
where the time-dependent field $\vec{h}_i(s)$ acting on spin $i$ is a sum of a decaying transverse field (along the $x$ direction) and a growing coupling term along $z$ (c.f. \eqnref{eq:TFIM}):
\begin{eqnarray}
\vec{h}_i(s) \equiv  (1-s)  \Gamma \hat{e}_x- s \sum_j J_{i j} M^z_j \hat{e}_z \ ,
\label{eq:O3}
\end{eqnarray}
where $T$ is again the total evolution time.
We note that such a state can be represented as a tensor network state (PEPS or MPS), which we discuss below, with bond dimension $M=1$.
As shown in Ref.~\onlinecite{Wang:2013tga},
the dynamics of this model are related to the O(2) model studied in Ref.~\onlinecite{Smolin:2013vpa}.

The initial condition is to have all spins aligned along the $x$ direction. To introduce some amount of randomness, we perturb the ideal initial condition, $\vec{M}_i(0)  =  (1,0,0)$, and rotate each spin by a small, random angle.
At the end of the annealing schedule, we extract the Ising variables $s_i$ as the sign of the $z$-component of the spin, $s_i = \mathrm{sign}(M_i^z)$.

We can introduce some local entanglement
by dividing the system into clusters of size $n$ and treating the quantum Hamiltonian exactly
on each cluster, while performing a mean-field decoupling for the inter-cluster terms.
In this scheme, the wavefunction of the whole system is approximated as the product state of cluster
wavefunctions, where entanglement within each cluster is fully taken into account but the individual
clusters are not entangled with each other.

\begin{figure}[tbp]
\centering
  \includegraphics[width=9cm]{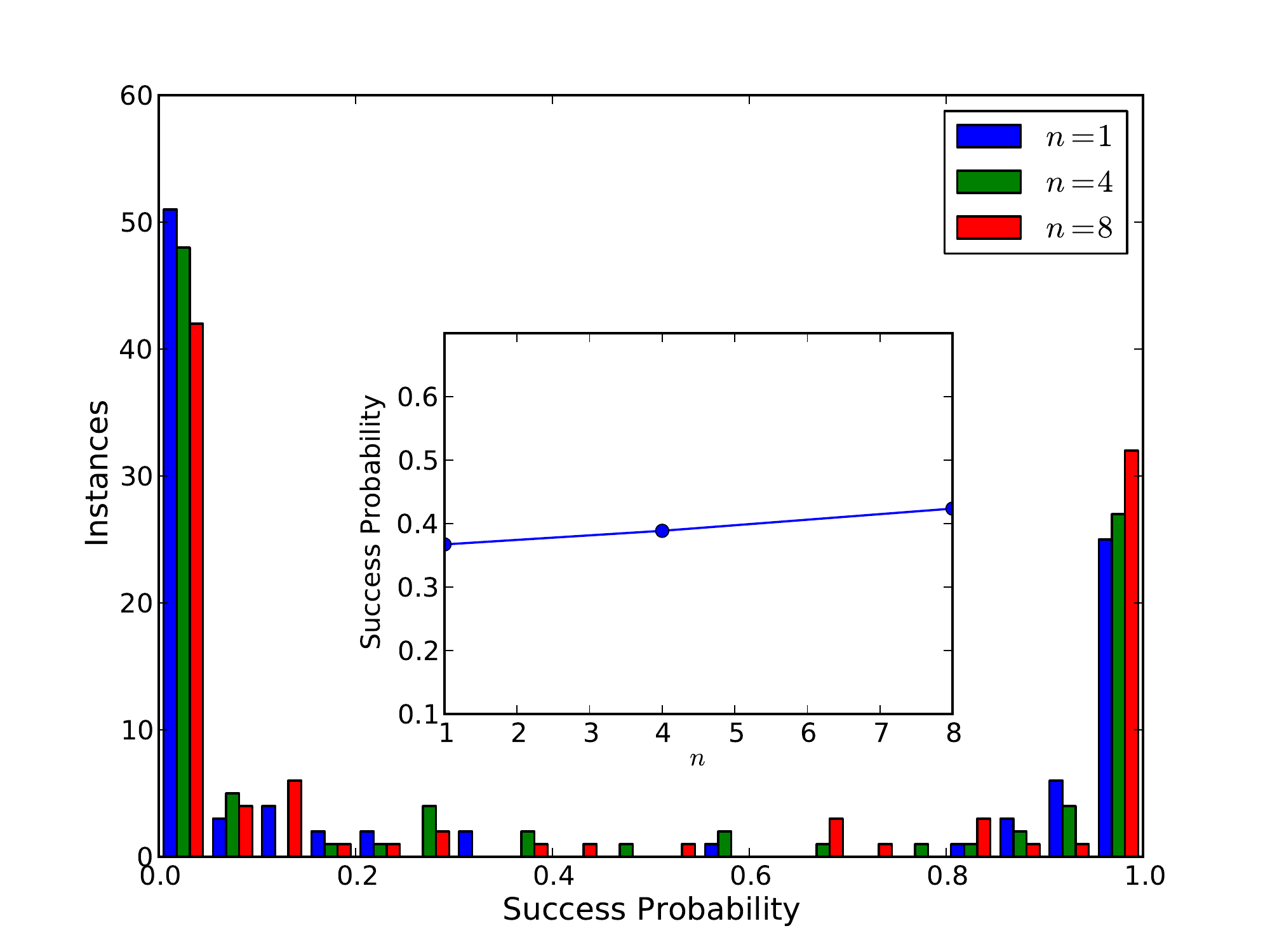}
\caption{The success probability histogram of the cluster mean-field annealing method. Inset shows the mean success probability for various cluster size $n$. }
\label{fig:cluster}
\end{figure}

To test this approach, we apply it to the same 1000 random instances on the 108-site chimera cluster as in Ref.~\onlinecite{Boixo:2014cg}.
We choose the cluster sizes to be $n=1,4,8$, and perform $1000$ trials for each instance with different initial states. Our results are shown
in Fig.~\ref{fig:cluster}.
Resolving the histogram of success probabilities, we observe that as the cluster size is increased, the number of instances that are solved with high probability increases, while the number of instances that cannot be solved decreases. In all cases, the distribution is highly bimodal, i.e. most instances are solved either all of the time or rarely regardless of the small variations in the initial condition.
Even when including entanglement on the 8-site clusters, the average success probability remains low, as shown in the inset of Fig.~\ref{fig:cluster}.

\section{Tensor network representations}

Another systematic way to include entanglement in the ansatz wavefunction is to
use a tensor network representation of the state. Tensor networks encompass several
classes of variational ansatz states that are constructed to efficiently describe weakly
entangled quantum states. Here, weakly entangled generally means that they obey an
area law, as is expected for the ground states of local Hamiltonians.
The most prominent and simplest example of a tensor network is the matrix product state (MPS), which forms the basis of the well-known density-matrix renormalization group method.~\cite{white1992} While the
MPS ansatz is most efficient in one dimension and generally faced with an exponential scaling in
higher dimensions, reliable numerical algorithms exist to perform simulations also
in higher dimensions or for systems with non-local interactions.
In recent years, a number of generalizations to higher dimensions have
been developed, particularly projected entangled-pair states (PEPS)\cite{verstraete2004} which hold great
promise to describe entangled states of two-dimensional quantum systems.

In all these states, a refinement parameter referred to as bond dimension $M$ 
regulates the amount of entanglement captured in the state. In the example of an MPS, the
von Neumann entanglement entropy $S$ between two parts of the system connected by a bond
of bond dimension $M$ is bounded by $S \leq \log M$. Note that $M=1$ corresponds to
the mean-field dynamics.

To simulate the system of \eqnref{eq:Ising} using an MPS, the system has to be mapped to a one-dimensional chain. If the original system has some spatial structure, different mappings may lead to very different results. For our test cases here, however, we are faced with a very non-local graph and thus expect very little difference between mappings. To simulate time evolution, we make use of the time-evolving block decimation (TEBD)\cite{vidal2003,daley2004} algorithm combined with a swap-operator approach\cite{stoudenmire2010} to deal with the non-local nature of interactions. The cost of a single timestep is thus $\mathcal{O}(N^2 M^3)$, where $N$ is the total number of spins. This, along with having to perform averaging over many different runs, limits the bond dimension we can reach to about $M \approx 30$.

\begin{figure}[tbp]
\centering
  \includegraphics{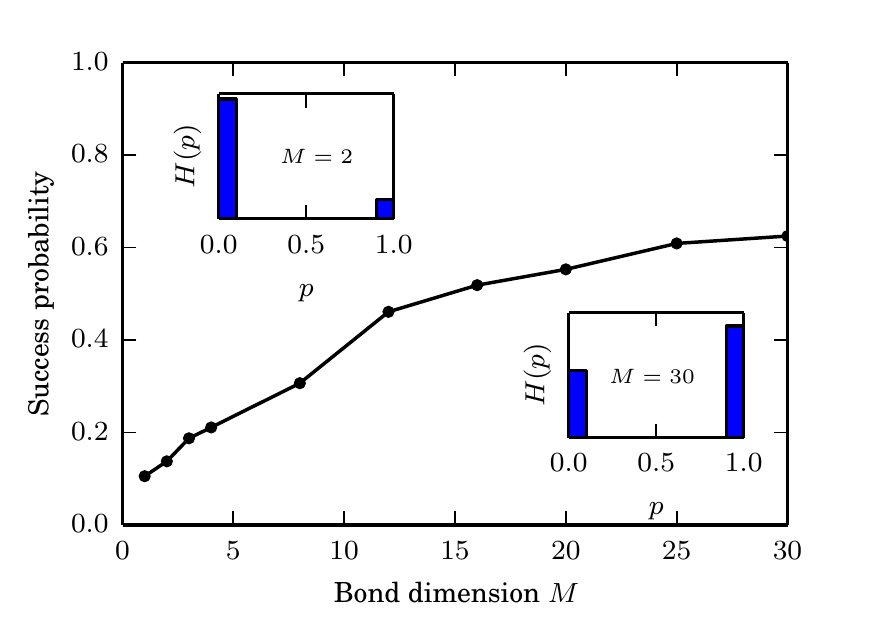}
  \caption{The average success probability versus the MPS bond dimension $M$ for the D-wave
  108-bit instances, with $T=50$ and $dt=0.2+0.1i$. Insets: Histogram of success probability $p$ for two values of $M$. }
\label{fig:M_vs_success}
\end{figure}

At the end of each annealing run, we collapse the matrix-product state to a single product state in the $Z$ basis
by sequentially projectively measuring $\sigma_i^z$ on each site, and compare the energy of this product state
against the known ground state energy of the spin glass problem.
The average success probability for the MPS approach is shown in Fig.~\ref{fig:M_vs_success}, where we
start from a fully polarized initial state, and the
average is performed over up to 1024 instances.
We observe that the success probability increases monotonically with the bond dimension $M$,
indicating that increasing the entanglement helps.
This is to be contrasted with the cluster mean-field approach in Sec.~\ref{sec:clustermf}, where inclusion of purely local
entanglement within a cluster has not significantly enhanced the probability of success.
Weak, yet long-ranged entanglement thus seems to be more helpful than
strong, but strictly local entanglement.

Similar to the mean-field approach, we can slightly perturb the initial state of the MPS evolution. As expected,
this improves the success probability for small bond dimension, but seems to be less helpful or even
detrimental for large bond dimension (not shown).

\begin{figure}[!t]
  \centering
  \includegraphics[width=3.5in]{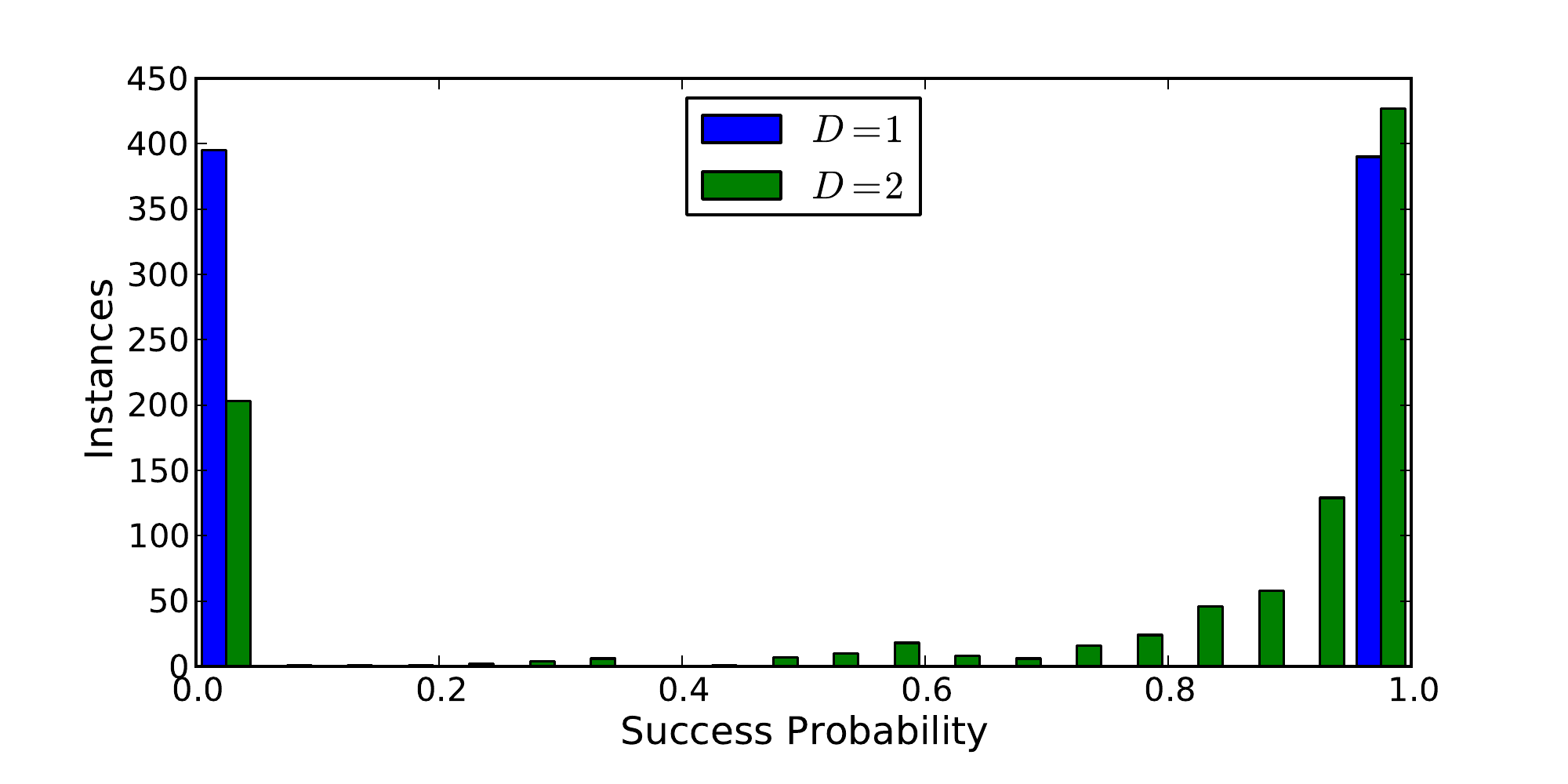}
  \caption{Histograms of lower bound success probability (see text for definition) from PEPS calculation.}
  \label{fig:peps}
\end{figure}

In a situation where the couplings
do have some additional structure, for example in a square lattice bilayer
system,\footnote{The bilayer system is chosen to ensure NP-hardness of the optimization problem.} it may be
advantageous to use an ansatz that is tailored to this structure. An example of such an ansatz
are PEPS,~\cite{verstraete2004} which can be thought of as the extension
of MPS to two-dimensional lattices. As such, they should be able to efficiently represent states
on two-dimensional lattices that exhibit an area law for the entanglement entropy, as opposed
to MPS which require an exponential number of states to do so.

We perform PEPS calculations for a square lattice bilayer system of $2 \times 10 \times 10$ sites with random $\pm 1$ couplings. We apply the algorithms for regular square lattice PEPS by grouping two sites, one from each layer, to a supersite with local dimension $d=4$. Due to the unfavorable scaling of the computational cost with the bond dimension $D$, we limit ourselves to bond dimensions $D=1,2$, and to annealing times of $T=1000$ for $D=1$ and $T=40$ for $D=2$. Similar to the MPS case, we use complex time evolution. Simulations are performed by starting from slightly perturbed initial states. At the end of the simulation, we measure the ground state energy $E_f$ of the final PEPS; assuming that this final state is a superposition of only the ground state and the first excited state, with energies $E_0$ and $E_1$, we can determine a lower bound for the square of the overlap with the ground state $p = |\langle \Psi | E_0 \rangle|^2$ using $E_f = p E_0 + (1-p) E_1$. In the following, we will discuss the average of the success rate $p$ over many different initial states.

In Fig.~\ref{fig:peps}, we show a histogram of $p$. We observe that for $D=1$, the histogram is strongly bimodal, and about half the instances appear to be easily solved (success probability $p > 0.9$), whereas the other half are never solved (success probability $p < 0.1$). This is consistent with the mean-field approach. For $D=2$, however, the picture changes drastically and the number of very hard instances with $p < 0.1$ is greatly suppressed, while the number of easy instances $p > 0.9$ is increased. We also observe that there are intermediate instances in particular in the regime $0.5 < p < 0.9$.

\section{Conclusions}

We have observed that compared to mean-field spin dynamics, even a modest amount of entanglement is a very useful resource for adiabatic quantum optimization.
With small bond dimensions of up to $M=30$ in an MPS simulation, or tiny bond dimensions of just $M=2$ in PEPS, the success probabilities of annealing to the ground
state of a non-planar spin glass problem are greatly enhanced.

While the potential for quantum annealing applied to real-world applications remains unclear, this might be taken as indication that imperfect qubits with
limited amount of entanglement may be useful as a computing platform. Conversely, approximating quantum annealing by efficient classical algorithms may provide
interesting classical optimization algorithms.

\acknowledgements

We acknowledge discussions with Dave Wecker and Michael Freedman.
This work has been supported by Microsoft Research and the Swiss National Science Foundation through the National Competence Center in Research QSIT.

\bibliography{quantumannealing}

\end{document}